\documentclass[aps,pra,reprint,superscriptaddress,10pt]{revtex4-2} 

\usepackage{amsmath,amssymb,graphicx,textcomp,gensymb,csquotes}

\begin{document}
\title{Kelvin's Chirality of Optical Beams}

\author{Sergey Nechayev}
\thanks{These two authors contributed equally}
\affiliation{Max Planck Institute for the Science of Light, Staudtstr. 2, D-91058 Erlangen, Germany}
\affiliation{Institute of Optics, Information and Photonics, University Erlangen-Nuremberg, Staudtstr. 7/B2, D-91058 Erlangen, Germany}
\affiliation{Department of Physics, University of Ottawa, 25 Templeton St., Ottawa, ON K1N\,6N5, Canada}
\affiliation{Max Planck—University of Ottawa Centre for Extreme and Quantum Photonics, 25 Templeton St., Ottawa, ON K1N\,6N5, Canada}

\author{J{\"o}rg S.~Eismann}
\thanks{These two authors contributed equally}
\affiliation{Max Planck Institute for the Science of Light, Staudtstr. 2, D-91058 Erlangen, Germany}
\affiliation{Institute of Optics, Information and Photonics, University Erlangen-Nuremberg, Staudtstr. 7/B2, D-91058 Erlangen, Germany}
\affiliation{Max Planck—University of Ottawa Centre for Extreme and Quantum Photonics, 25 Templeton St., Ottawa, ON K1N\,6N5, Canada}
\affiliation{Institute of Physics, University of Graz, NAWI Graz, Universit\"atsplatz 5, Graz, 8010 Austria}

\author{Rasoul Alaee}
\affiliation{Department of Physics, University of Ottawa, 25 Templeton St., Ottawa, ON K1N\,6N5, Canada}
\affiliation{Max Planck—University of Ottawa Centre for Extreme and Quantum Photonics, 25 Templeton St., Ottawa, ON K1N\,6N5, Canada}

\author{Ebrahim Karimi}
\affiliation{Max Planck Institute for the Science of Light, Staudtstr. 2, D-91058 Erlangen, Germany}
\affiliation{Department of Physics, University of Ottawa, 25 Templeton St., Ottawa, ON K1N\,6N5, Canada}
\affiliation{Max Planck—University of Ottawa Centre for Extreme and Quantum Photonics, 25 Templeton St., Ottawa, ON K1N\,6N5, Canada}

\author{Robert W.~Boyd}
\affiliation{Department of Physics, University of Ottawa, 25 Templeton St., Ottawa, ON K1N\,6N5, Canada}
\affiliation{Max Planck—University of Ottawa Centre for Extreme and Quantum Photonics, 25 Templeton St., Ottawa, ON K1N\,6N5, Canada}

\author{Peter Banzer}
\email[]{peter.banzer@uni-graz.at}
\affiliation{Max Planck Institute for the Science of Light, Staudtstr. 2, D-91058 Erlangen, Germany}
\affiliation{Institute of Optics, Information and Photonics, University Erlangen-Nuremberg, Staudtstr. 7/B2, D-91058 Erlangen, Germany}
\affiliation{Max Planck—University of Ottawa Centre for Extreme and Quantum Photonics, 25 Templeton St., Ottawa, ON K1N\,6N5, Canada}
\affiliation{Institute of Physics, University of Graz, NAWI Graz, Universit\"atsplatz 5, Graz, 8010 Austria}


\date{\today}
\begin{abstract}
Geometrical chirality is a property of objects that describes three-dimensional mirror-symmetry violation and therefore it requires a non-vanishing spatial extent. In contrary, optical chirality describes only the local handedness of electromagnetic fields and neglects the spatial geometrical structure of optical beams. In this manuscript, we put forward the physical significance of geometrical chirality of spatial structure of optical beams, which we term \textit{Kelvin's} chirality. Further, we report on an experiment revealing the coupling of Kelvin's chirality to optical chirality upon transmission of a focused beam through a planar medium. Our work emphasizes the importance of Kelvin's chirality in all light-matter interaction experiments involving structured light beams with spatially inhomogeneous phase and polarization distributions.
\end{abstract}

\maketitle 

\section{Introduction}
Since its first definition by Lord Kelvin in 1893 \cite{Kelvin1894}, the term “chiral” has found its use across the fields of physics, mathematics, chemistry and biology. Chirality describes mirror-symmetry violation --- if an object cannot be superimposed with its own mirror image by means of rotations and translations, it is termed chiral~\cite{Kelvin1894,Barron1986,Barron1986_01_01,BarronCosmic2012}. Consequently, \textit{geometrical} chirality is inherently a non-local three-dimensional (3D) structural property of objects that requires a non-vanishing spatial extent~\cite{Buda1992,Buda1992a,Rassat2004,Fowler2005,Efrati2014,Ivan_chirality_obj2016,Gutsche2020}.
On the other hand, \textit{optical} chirality is a bit less tangible. Using parity inversion transform $\hat{\text{P}}$ (a point reflection)~\cite{Kelvin1894,Barron1986,Barron1986_01_01,BarronCosmic2012} to define optical chirality allows electromagnetic beams to be chiral at specific points in space. The common definition of optical chirality as $C \propto \text{Im} ( \textbf{E} \cdot \textbf{H}^\ast )$~\cite{Tang2010,Bliokh2011_02,Chirality_andrews_2012,AndrewsMeasures12,Bliokh2014_07,Nieto2017,cameron_robert_p._chirality_2017,Crimin2019,Poulikakos2019,Mackinnon2019}, where \textbf{E} and \textbf{H} refer to the electric and magnetic field vectors, respectively, is in perfect agreement with this.\\
%
%
Optical chirality plays a crucial role in chiral light-matter interactions~\cite{Tang2010,valev_chirality_2013,Bliokh2014_07,Signatures_chirality_andrews2015,Lodahl2017}. Tools such as circular dichroism spectroscopy are widely used for distinguishing molecular enantiomers, studying proteins' structure and measuring the composition of materials \cite{Woody1995, barron_2004}.
%
In all these scenarios, chiroptical phenomena are used to determine whether the interacting matter features geometrical chirality originating from its spatial extent.\\
However, polarization and phase distributions of optical beams can exhibit fascinating topological features in 3D space as well. These include knotted and linked polarization and phase singularities~\cite{Berry2001,Leach2004,Larocque2018} and even polarization M\"obius strips~\cite{Bauer2015}. Some of these peculiar field topologies are asymmetric upon parity transformation, rendering the optical beams \textit{geometrically} chiral. Nonetheless, these beams do not necessarily also exhibit optical chirality. Since the definition of optical chirality $C$ only refers to a local arrangement of \textbf{E} and \textbf{H}, it fails to describe any form of chirality originating from the spatial extent of the beams. This geometrical chirality of the spatial polarization- and phase-structure of optical beams (hereinafter referred to as \textit{Kelvin’s} chirality or $K$) may be directly involved in chiral light-matter interaction. Indeed, the corkscrew wavefronts of linearly polarized Laguerre-Gaussian beams $(C=0)$ readily engage in chiral light-matter interactions \cite{Toyoda2012,Wong446,Gorodetski2013_farfieldChirality,Forbes2018, nechayev_orbital--spin_2019, wozniak_interaction_2019,Andrews2019a,Andrews2019b}. Also the polarization structure of vector beams with $C=0$ can produce circularly polarized light (CPL) upon scattering by achiral particles~\cite{Eismann2018,Mimicking}. Kelvin's chirality therefore represents an electromagnetic equivalent of geometrical chirality and refers to a certain chiral arrangement of electromagnetic fields in 3D space. It complies with the point-reflection geometrical non-local definition of chirality, but it can not be described involving optical chirality $C$.\\
To explore the phenomenon of Kelvin’s chirality, we construct an experiment, where a normally incident polarization-structured cylindrically symmetric beam with $C=0$ and $K \neq 0$ is focused and transmitted through a planar stratified medium. Surprisingly, we observe CPL in transmission ($C\neq 0$) with its handedness being dependent on the incident $K$. We elucidate our results using a simple geometrical model, generalized Fresnel coefficients of the layered sample, and helicity conservation laws~\cite{Ivan2012,Cameron_helicity2012,cameron_robert_p._chirality_2017,barnett_duplex2012,Ivan2013,Bliokh2013,Nieto2015,Nieto2017,Nieto2017_05,ivan_unified2017,Crimin2019,Poulikakos2019,Mackinnon2019}. Our work calls for a careful evaluation of Kelvin's chirality as a crucial component of all light-matter interaction experiments that involve light beams with spatially inhomogeneous phase or polarization distributions.
\section{Theory} \label{chap_theo}
\begin{figure}[htbp]
\centerline{\includegraphics[width=1\columnwidth]{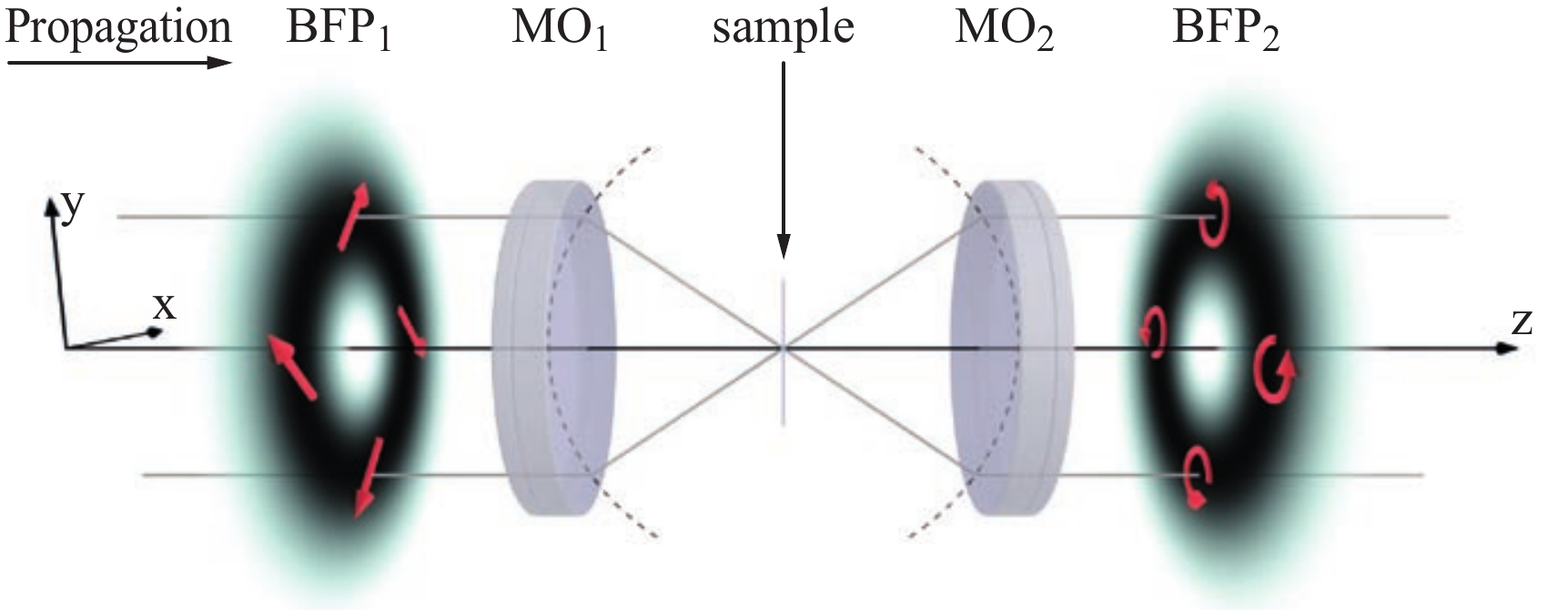}}
\caption{Schematic illustration of the system. The incoming beam, propagating from left to right, is focused and collimated by two confocally aligned aplanatic microscope objectives (MOs). The sample is positioned normally to the $z$-axis between the MOs. The red arrows depict the local polarization of the beam before and after the introduction of a relative phase of $\pi/2$ between radial and azimuthal polarization components.}
\label{fig_sketch}
\end{figure}	
We investigate a system of two dry aplanatic microscope objectives (MO$_1$ and MO$_2$) in confocal alignment, as sketched in Figure \ref{fig_sketch}, with an incident monochromatic beam propagating along the $z$-axis. Focusing by an aplanatic MO$_1$ converts the spatial coordinates $(x_1,y_1)$ in the back focal plane (BFP$_1$) of MO$_1$ to the angular coordinates of the focused beam via $k_x=-x k / f$, $k_y = -y k / f$, where $f$ is the focal length, $k=2\pi / \lambda$ is the wavenumber and $\lambda$ is the wavelength~\cite{novotny_principles_2012}. Collimation by MO$_2$ induces a reverse transformation such that in BFP$_2$ we obtain $(x_2,y_2)=(k_x f / k, k_y f /k)$. To describe the evolution of polarization pattern, we decompose the incident beam into radial and azimuthal polarization components, which correspond to the transverse magnetic (TM or $p$) and transverse electric (TE or $s$) polarization components of the focused beam, respectively. Reference~\cite{novotny_principles_2012} describes the process in detail.\\
Consider now a spirally polarized incident vector beam (see Fig.~\ref{fig_sketch}), which is an in- or $\pi$-out-of-phase superposition of a radially and an azimuthally polarized mode~\cite{spiral_beam,Eismann2018,Mimicking,Nechayev2020}. The electric field in the BFP$_1$, which we assume to coincide with the beam's waist position ($z=0$), can be written as:
\begin{align}
&\mathbf{E}^{\text{in}}_\sigma =E\left(\rho\right)    \left[ \hat{\boldsymbol{\rho}} +\sigma \hat{\boldsymbol{\varphi}}\right]
\text{,}
\label{eq:incident_beam} \end{align}
where \mbox{$E\left(\rho\right)=E_0\frac{\rho}{w_0}\exp\left(- \frac{\rho^2}{w_0^2}\right) $}, with $w_0$ being the beam waist, $E_0$ is a constant, $\rho$ is the radial cylindrical coordinate, $\hat{\boldsymbol{\rho}}$ and $\hat{\boldsymbol{\varphi}}$ are the radial and axial unit vectors, respectively, and $\sigma = \pm 1$, corresponding to an in- or $\pi$-out-of-phase superposition. %
We note that parity transformation applied to the beam in Eq. (\ref{eq:incident_beam}) inverses the direction of the spiral polarization $\hat{\text{P}}  \left\{ \mathbf{E}^{\text{in}}_\sigma \right\} = \mathbf{E}^{\text{in}}_{-\sigma}$ \cite{Mimicking}. Therefore, without assigning an exact value to $K$, we can argue that the beam in Eq. (\ref{eq:incident_beam}) possesses a non-zero Kelvin's chirality $K(\mathbf{E}^{\text{in}}_{\pm\sigma})\neq0$ and that a pair beams with $\sigma=\pm 1$ constitute a pair of chiral enantiomers, which implies that $K(\mathbf{E}^{\text{in}}_{+\sigma})=-K(\mathbf{E}^{\text{in}}_{-\sigma})$.
 At the same time, such beams feature zero optical chirality $C$, which in the paraxial regime can be expressed as the third Stokes parameter $S_3$~\cite{Tang2010,Bliokh2011_02,Chirality_andrews_2012,AndrewsMeasures12,Nieto2017,cameron_robert_p._chirality_2017,Crimin2019,Poulikakos2019,Mackinnon2019}:
\begin{align} 
C^{\text{in}}(\rho) \propto S_3^{\text{in}}(\rho) = 2\text{Im} \left\{\left[\mathbf{E}^{\text{in}}_\sigma \cdot  \hat{\boldsymbol{\rho}}\right]^\ast \cdot \left[\mathbf{E}^{\text{in}}_\sigma \cdot  \hat{\boldsymbol{\varphi}}\right] \right\} \equiv 0
\text{.}\label{eq:incident_chirality} \end{align}
We introduce a planar layered structure between the MOs (Fig. \ref{fig_sketch}) positioned normally to the $z$-axis. Transmission through a stratified medium strongly depends on the polarization ($p,\,s$) and angular coordinates ($k_x,\,k_y$) of the focused beam. The angle of incidence of each angular component ($k_x,\,k_y$) of the focused beam is defined by $\sin^{-1}\left(k_\rho \right/k)$, where $k_\rho=\sqrt{k_x^2+k_y^2}$~\cite{novotny_principles_2012}.
 Our sample is designed such that the Fresnel coefficients $t_p(k_\rho)$ and $t_s(k_\rho)$ for the transmitted $p$- and $s$-polarized field components at an angle of $30\degree$ with respect to the surface normal acquire a $\pi/2$ phase difference, but have equal amplitudes $t_p/t_s \approx \exp (\imath \pi/2)$~\cite{Yeh77}. As a result, the transmitted field $\mathbf{E}^{\text{tr}}_\sigma $, the intensity $S_0^{\text{tr}}$ and the third Stokes parameters $S_3^{\text{tr}}$ are:
\begin{align}
\begin{split} 
&\mathbf{E}^{\text{tr}}_\sigma(\rho) =E(\rho)    \left[ t_p(\rho)\hat{\boldsymbol{\rho}} +\sigma t_s(\rho)\hat{\boldsymbol{\varphi}} \right]\text{,}\\
&S_0^{\text{tr}}(\rho) = \left|E (\rho) \right|^2 \left[ \left| t_p(\rho)  \right|^2+ \left| t_s (\rho) \right|^2   \right] \text{,} \\
&S_3^{\text{tr}}(\rho) =  -2\sigma \left|E (\rho) \right|^2 \left|t_p(\rho) t_s(\rho) \right|
\text{,}
\label{eq:transmitted_beam} \end{split}\end{align}
where $\rho = \sqrt{x_2^2+y_2^2} =f k_\rho /k $ and the angle of $30\degree$ is given by the condition $\rho=0.5f$ in BFP$_1$. Equations (\ref{eq:transmitted_beam}) show that the transmitted beam at this angle is circularly polarized $S_3^{\text{tr}}/S_0^{\text{tr}}\approx -\sigma$ with the handedness being dependent on the spatial polarization distribution of the incident beam. This proves that Kelvin's chirality of the incident beam $K(\sigma)$ can perfectly couple to optical chirality $C$ of the transmitted beam in a simple cylindrically symmetric scenario.
 
%
%
%
%
%
%
\section{Experiment}
\begin{figure}[htbp]
\centerline{\includegraphics[width=1\columnwidth]{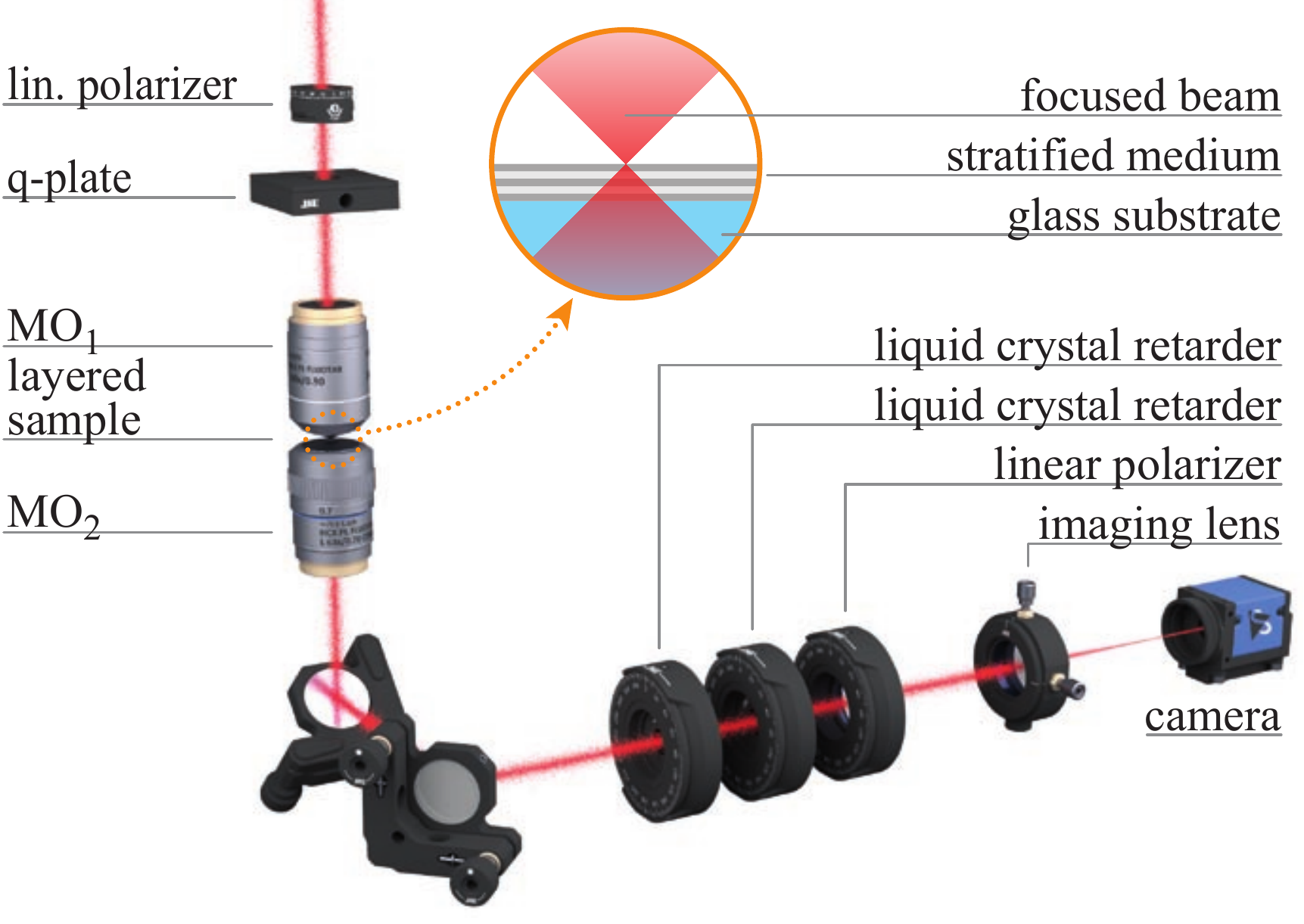}}
\caption{Experimental setup. A linear polarizer and a q-plate of charge 1/2 convert the incoming Gaussian beam into the desired mode. Two confocally aligned microscope objectives focus the incoming beam onto the sample and collimate it subsequently. Two liquid crystal variable retarders, a linear polarizer and a lens are used to perform a polarization resolved imaging of the back focal plane of the second microscope onto a camera.}
\label{fig_setup}
\end{figure}
\begin{figure*}[htbp]
\centerline{\includegraphics[width=2\columnwidth]{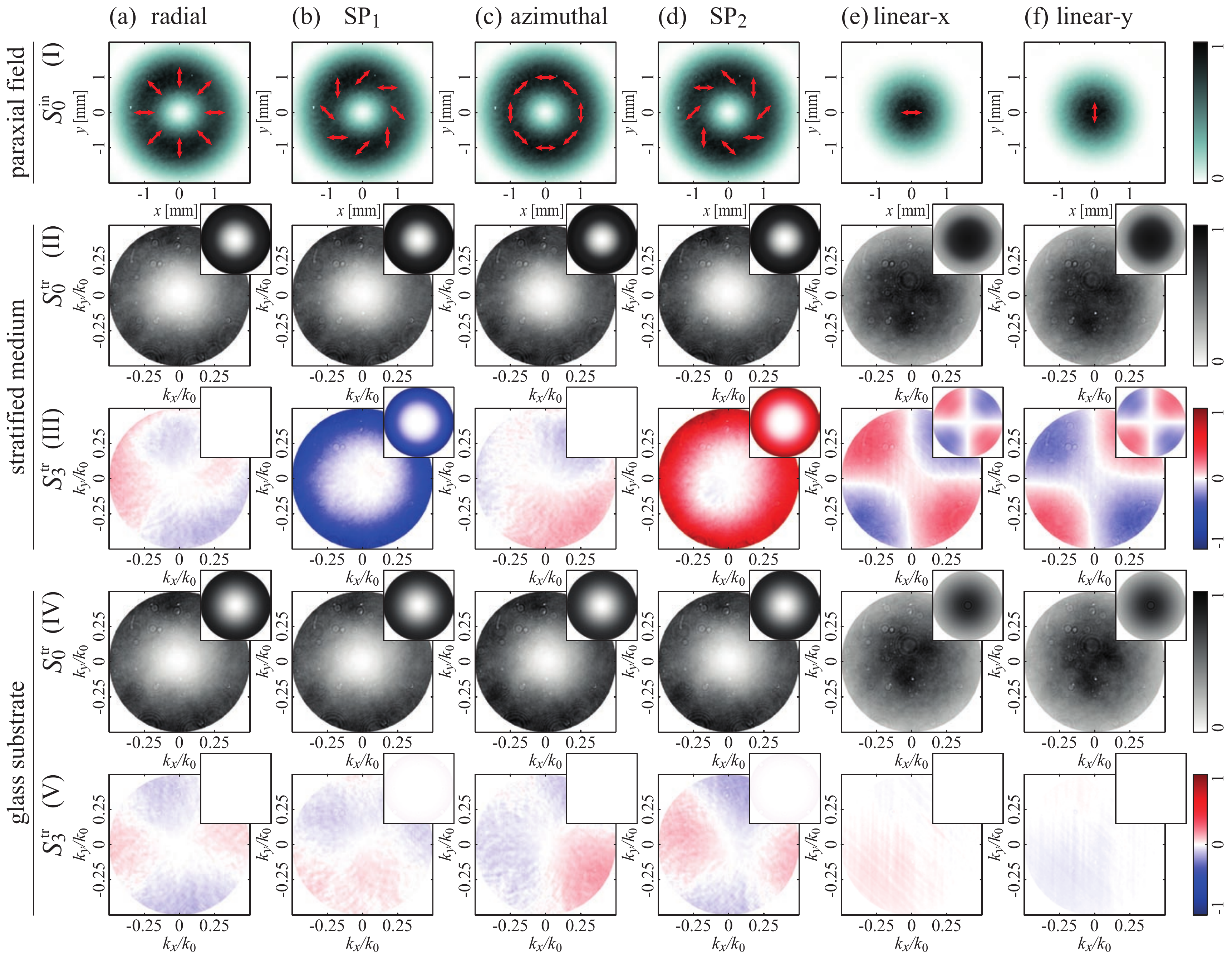}}
\caption{Measurements for six input polarizations: (a) radial, (b) spiral SP$_1$, (c) azimuthal, (d) spiral SP$_2$, (e) linear-$x$, (f) linear-$y$. (I) Intensity distributions before entering the first microscope objective. The local polarization state is indicated by red arrows. Stokes parameters $S_0^{\text{tr}}$ (II) and $S_3^{\text{tr}}$ (III), measured in transmission through the stratified medium. Theoretical counterparts are plotted as insets. $S_0^{\text{tr}}$(IV) and $S_3^{\text{tr}}$(V) correspond to transmission through a glass substrate.}
\label{fig_results}
\end{figure*}

	
Figure \ref{fig_setup} illustrates our experimental setup~\cite{Banzer2010}. We transmit a Gaussian beam (wavelength $\lambda$ = 532\,nm, linewidth $\Delta\lambda_{\text{FWHM}} \approx$ 4\,nm) through a linear polarizer (LP) and a q-plate of charge $1/2$~\cite{Marrucci2006,Larocque_2016}. Depending on the relative angle between the LP and the q-plate, this results in a radially, azimuthally or one of two spirally polarized doughnut-shaped beams SP$_1$ $(\sigma = + 1)$ and SP$_2$ $(\sigma = - 1)$. For comparison, we also performed the measurements for linear $x$- and $y$-polarized input beams. 
The investigated stratified medium on top of a glass substrate was fabricated by \textit{Iridian Spectral Technologies} such that $t_p/t_s \approx \exp (\imath \pi/2)$ at the angle of incidence of $30\degree$ for the wavelength of 532\,nm. The numerical aptertures (NA) of the two MOs were chosen to be larger than 0.5 to cover this angle.
Additionally, we repeated the measurements on a bare glass substrate with refractive index of $n=1.52$ and thickness of $170 \,\mu\mathrm{m}$. Since the coherence length of our laser is only $ \lambda^2/(n \Delta \lambda) \approx 47 \,\mu\mathrm{m}$, such a measurement provides reference data for the case where the sample does not introduce a phase shift between $p$ and $s$-polarizations~\cite{Yeh77}.
We use two liquid crystal variable retarders and an LP to project the transmitted beam onto an arbitrary polarization state. Finally, we image the polarization resolved intensity distribution in the BFP$_2$ onto a CMOS camera for full spatial Stokes vector reconstruction.\\
First, Figure \ref{fig_results} (I) shows the measured input intensity distributions $S_0^\text{in}= \left[I_\text{RCP}^\text{in}+I_\text{LCP}^\text{in}\right]$, where $I_\text{RCP}$ and $I_\text{LCP}$ are the measured input intensities of the RCP and LCP polarizations. The red arrows depict the corresponding polarization patterns. Secondly, in Figure \ref{fig_results} (II) and (III) we plot the intensity distribution $S_0^{\text{tr}}$ and the normalized third Stokes parameters $S_3^{\text{tr}} = \left[I_\text{RCP}-I_\text{LCP}\right]/max(S_0^{\text{tr}})$ of the light transmitted through the stratified sample. The insets show the theoretical predictions. As expected, the radial and azimuthal beams do not generate CPL, while both spiral beams SP$_1$ and SP$_2$ strongly couple to CPL with the sign of the transmitted $S_3^{\text{tr}}$ depending on the spiral orientation of the incoming beam or, equivalently on the sign of Kelvin's chriality $K(\sigma)$ (see Fig. \ref{fig_results} (III) (a)-(d)). In fact, for the angle of $30 \degree$ ($\text{NA}=0.5$) we achieve almost perfect coupling to CPL with values of $S_3^{\text{tr}}$ approaching $\mp 1 = -\sigma$, as expected from Eqs. (\ref{eq:transmitted_beam}). For linear input polarizations, the cylindrical symmetry of the beam is broken, resulting in a four-fold pattern of CPL in Figure \ref{fig_results} (III) (e) and (f). Nevertheless, since linearly polarized beams do not possess any Kelvin's chirality, their average generated $S_3^{\text{tr}}$ is zero. Finally, in Figure \ref{fig_results} (IV) and (V) we present the transmission measurements through the glass sample. Here, in Figure \ref{fig_results} (V) (a)-(d) we observe only a residual pattern of $S_3^{\text{tr}}$ with zero average, resembling the measurements for the radial and azimuthal polarization through the stratified medium (III) or the glass sample (V) (a) and (c). This resemblance indicates that the residual $S_3^{\text{tr}}$ originates from the imperfections of the q-plate, as confirmed by the measurements of linear polarizations transmitted through the glass substrate in Figure \ref{fig_results} (V) (e) and (f), which we performed without the q-plate.
\section{Discussion and Conclusion}
To comprehend the role of Kelvins’ chirality it is worth discussing our result from the point of view of geometry, material composition and conservation laws.\\
First, in chiral light-matter interactions it is not exclusively the chirality of matter that couples to optical chirality, but essentially the geometrical chirality of the whole experiment. For instance, structurally achiral planar metamolecules show chiroptical response at normal incidence, if the mirror symmetry is broken by the heterogeneous material composition of their constituents~\cite{Banzer2016,NechayevTrimer2019}. Additionally, chiroptical effects also appear if a planar structure and the k-vector of an obliquely incident CPL form a geometrically chiral arrangement~\cite{Bunn1945}.
%
%
Similarly, linear dipoles can emit chiral light, when appropriately positioned close to an optical waveguide~\cite{JoosDipole2018,Neugebauereaav7588}. In this regard, Kelvins’ chirality ensures $\hat{\text{P}}$-symmetry breaking at the level of beam geometry.\\
Second, we may ask which physical conservation laws permit generation of CPL in a cylindrically symmetric system? Helicity \--- the projection of spin on the propagation direction \--- characterizes the “handedness” of a beam~\cite{Ivan2012,Cameron_helicity2012,cameron_robert_p._chirality_2017,barnett_duplex2012,Ivan2013,Bliokh2013,Nieto2015,Nieto2017,Nieto2017_05,ivan_unified2017,Crimin2019,Poulikakos2019,Mackinnon2019}. Helicity is only preserved in electromagnetically dual (impedance matched) conditions, equivalent to $t_p=t_s$ at all angles for planar systems~\cite{Ivan2012,Ivan2013}. Neither the stratified medium nor the glass substrate preserve helicity. Our focused beam acquires circular polarization upon transmission through the stratified medium for the same physical reason that obliquely incident plane-wave CPL at an air-glass interface acquires elliptical polarization in ellipsometry measurements, i.e, the difference of the two Fresnel coefficients.\\
Third, we construct the spiral beams with Kelvin’s chirality by a superposition of a radially and an azimuthally polarized beams. Neither of these beams alone possesses Kelvin’s chirality. However, the radial and azimuthal beams are $\hat{\text{P}}$-even and $\hat{\text{P}}$-odd, respectively, relative to a reflection plane that contains the $z$-axis, which breaks the $\hat{\text{P}}$-symmetry of their superposition. Surprisingly, a $\hat{\text{P}}$-even radial beam does not violate the parity odd transformation of $\textbf{E}$ itself, which holds at each point of space such that $\hat{\text{P}} \left\{\textbf{E} \right\}= -\textbf{E}$ (see Fig. 1(c) in~\cite{Mimicking}). Moreover, we consider the geometry of spiral beams along with their direction of propagation, which renders them asymmetric under time reversal ($\hat{\text{T}}$) and symmetric under combined $\hat{\text{P}}\hat{\text{T}}$ inversion. Previously, L.D. Barron defined these transformation properties as "false chirality" in molecular systems~\cite{Barron1986,Barron1986_01_01,BarronCosmic2012} and they are still a subject of active research~\cite{Bliokh2014_07}. At the same time, corkscrew wavefronts of linearly polarized Laguerre-Gaussian beams are asymmetric under $\hat{\text{P}}$ and symmetric under $\hat{\text{T}}$ reversal, respectively, rendering them \enquote{truly chiral}. We envision that further classification of Kelvin’s chirality and transformation properties of structured beams may be necessary.\\
Finally, Kelvin’s chirality of optical beams, similarly to chirality of matter, is a function of a geometrical shape, which is independent of its optical manifestations and does not rely on any measurable observable. For instance, an unfocused spiral beam transmitted through the stratified sample would not acquire optical chirality since $t_p=t_s$ at normal incidence. Quantification of Kelvin’s chirality is therefore as elusive a task as quantification of chirality of matter, for which many attempts have been made, but no universal measure has been established to date~\cite{Buda1992,Buda1992a,Rassat2004,Fowler2005,Efrati2014,Ivan_chirality_obj2016,Gutsche2020}.\\
In conclusion, we presented an experiment where the geometrical chirality of an optical beam, termed here as Kelvin's chirality and manifested as parity asymmetric spatial polarization distribution, couples to optical chirality upon transmission of a focused beam through a planar medium in a cylindrically symmetric scenario. We elucidated the underlying mechanism of chiral light-matter interaction by symmetry, material composition, helicity conservation laws and a simple analytical model. Our results emphasize that spatially inhomogeneous phase and polarisation profiles of structured light beams constitute an important degree of freedom in chiral light-matter interactions beyond optical chirality.
%
%
%
\section*{Acknowledgments}
R.A. acknowledges the support of the Alexander von Humboldt Foundation through the Feodor Lynen Fellowship. R.A. and R.W.B. acknowledge support through the Natural Sciences and Engineering Research Council of Canada,the Canada Research Chairs program, and the Canada First Research Excellence Fund.
\bibliographystyle{apsrev4-1}
\bibliography{bib}


\end{document}